\documentclass{article}

\usepackage[utf8]{inputenc}
\usepackage[margin=1in]{geometry}
\usepackage{xcolor}

\usepackage{color,soul}
\usepackage{authblk}
\usepackage{graphicx}
\usepackage{hyperref}
\usepackage{tabularx}
\usepackage{amsmath}
\usepackage{float}

\usepackage[numbers, square]{natbib}
\title{Optical and spin coherence of Er$^{3+}$ in epitaxial CeO$_2$ on silicon}
\author[1,2,*]{Jiefei Zhang}
\author[3,1]{Gregory D. Grant}
\author[3,1]{Ignas Masiulionis}
\author[3,1,2]{Michael T. Solomon}
\author[4]{Jasleen K. Bindra}
\author[4]{Jens Niklas}
\author[5,6,2]{Alan M. Dibos}
\author[4]{Oleg G. Poluektov}
\author[1,2,3]{F. Joseph Heremans}
\author[3,1,2,*]{Supratik Guha} 
\author[3,1,2,7]{David D. Awschalom}
\affil[1]{Materials Science Division, Argonne National Laboratory, Lemont, Illinois 60439, USA}
\affil[2]{Center for Molecular Engineering, Argonne National Laboratory, Lemont, IL 60439, USA}
\affil[3]{Pritzker School of Molecular Engineering, University of Chicago, Chicago, Illinois 60637, USA}
\affil[4]{Chemical Sciences and Engineering Division, Argonne National Laboratory, Lemont, IL 60439, USA}
\affil[5]{Nanoscience and Technology Division, Argonne National Laboratory, Lemont, Illinois 60439, USA}
\affil[6]{Center for Nanoscale Materials, Argonne National Laboratory, Lemont, Illinois 60439, USA}
\affil[7]{Department of Physics, University of Chicago, Chicago, Illinois 60637, USA}
\affil[*]{emails:jfzhang@anl.gov; sguha@anl.gov}

\date{\today}
\begin{document}
\maketitle



\section*{Abstract}

Solid-state atomic defects with optical transitions in the telecommunication bands, potentially in a nuclear spin free environment, are important for applications in fiber-based quantum networks. Erbium ions doped in CeO$_2$ offer such a desired combination. Here we report on the optical homogeneous linewidth and electron spin coherence of Er$^{3+}$ ions doped in CeO$_2$ epitaxial film grown on a Si(111) substrate. The long-lived optical transition near 1530 nm in the environmentally-protected 4f shell of Er$^{3+}$ shows a narrow homogeneous linewidth of 440 kHz with an optical coherence time of 0.72 $\mu s$ at 3.6 K. The reduced nuclear spin noise in the host allows for Er$^{3+}$ electron spin polarization at 3.6 K, yielding an electron spin coherence of 0.66 $\mu s$ (in the isolated ion limit) and a spin relaxation of 2.5 ms. These findings indicate the potential of Er$^{3+}$:CeO$_2$ film as a valuable platform for quantum networks and communication applications. 


\section{Introduction}

Rare-earth (RE) ions in dielectric solid-state hosts provide a promising platform for developing quantum memories \cite{Awschalom2018,Lvovsky2009,Heshami2016} in quantum repeaters \cite{Awschalom2018,Sangouard2011} for use in quantum communication networks because of their unique combination of stationary matter qubits and flying photon qubits provided by transitions in their uniquely environmentally protected 4f shell electrons. This spin-photon interface \cite{Awschalom2018,Wolfowicz2021} is characterized by long-lived spin states with a long coherence time \cite{Wolfowicz2021} and narrow optical homogeneous linewidth \cite{Kinos2021,Ourari2023}. These unique properties of RE ions have been explored to demonstrate quantum memories in atomic vapor using the DLZC (Duan, Lukin, Cirac and Zoller) protocol, \cite{Duan2001} and in solids, using electromagnetically induced transparency \cite{Lvovsky2009,Heshami2016}, photon-echo \cite{Lvovsky2009,Heshami2016} and atomic frequency comb \cite{Craiciu2019,Dutta2023}. Progress has also been made in realizing entanglement distribution \cite{Lago-Rivera2021} and quantum transduction \cite{Bartholomew2020,OBrien2014}. All these demonstrations are enabled by a combination of RE ion and crystalline host properties.

The implementation of quantum networks for quantum communication applications demands the realization of entanglement distribution of quantum information over long distances. Thus, it is desirable for the matter qubits/nodes to be able to interface with telecommunication C-band photons to leverage the existing optical fiber networks providing minimum optical loss. Trivalent erbium ions (Er$^{3+}$) embedded in rare-earth oxides have $^4I_{13/2}$ to $^4I_{15/2}$ optical transitions in the telecom C-band and, thus, have gained attention as a candidate system aimed at developing telecom-compatible quantum memories needed in quantum repeaters. Approaches have been explored to store information in Er$^{3+}$ long-lived optical transitions using photon-echo techniques \cite{Lvovsky2009,Sangouard2011,Lauritzen2010} with retrieval efficiencies up to 40\%, \cite{Dajczgewand2014} but with fidelities of recalled states well below the classical limit of 1/2 and no-cloning limit of 2/3 \cite{Huo2018}. These figures-of-merit are critical to prevent eavesdropping in quantum communication networks and enable acceptable levels of quantum error correction in distributed fault-tolerant quantum computing. Therefore, long storage times and efficient retrieval of states with high fidelity are essential for quantum memories. The optical storage times are limited by the optical coherence time, typically less than 1 $\mu$s \cite{Lvovsky2009,Sangouard2011,Lauritzen2010}, below the proposed requirement for quantum repeater based long-distance quantum networks \cite{Razavi2009}. The use of Er$^{3+}$ spin states has emerged as a promising alternative to store photon information with much longer storage times, \cite{Rancic2018,Probst2015} where a collective spin relaxation in the atomic ensemble is used as a local memory and reconverted to a photon through a collective interference effect \cite{Sangouard2011}. Therefore, it is critical to realize long-lived spin states with coherence times orders of magnitude longer than the optical excited state lifetime for efficient optical control of the spin state and for subsequently suitable long-term quantum state storage \cite{Wolfowicz2021,Rancic2018}. 

As a Kramer's ion, the intrinsic non-zero electronic magnetic moment of Er$^{3+}$ poses an intrinsic limit on the electronic spin relaxation and, therefore, the spin coherence. In addition, the presence of fluctuating magnetic field noise induced by the intrinsic electronic and nuclear spins in host materials further reduces the spin coherence \cite{Chirolli2008}. In some spin qubit-host systems, approaches have been to taken to engineer the nuclear spin bath density through isotopic purification of the host material \cite{Wolfowicz2021,Anderson2021,Bourassa2020,Balasubramanian2009}, reduction of unintended spin defects during synthesis, and optimization of defect creation \cite{Wolfowicz2021} and even isotopic doping processes using $^{167}$Er \cite{Tawara2017} to improve upon coherence properties of the targeted spin qubits. Alternatively, finding host materials with low natural abundant isotopes of non-zero nuclear spins is another viable pathway towards improving spin properties by minimizing spin noise in the host for network applications  \cite{Wolfowicz2021,kanai2022generalized}. An electron spin coherence time of 23 ms has been reported for Er$^{3+}$ at 10 mK in CaWO$_4$, which has a low nuclear spin environment with only 14\% of $^{183}$W isotope of natural abundant tungsten with nuclear spin of $I=\frac{1}{2}$ contributing to the spin noise in the host \cite{LeDantec2021CaWO4}. 

To this end, cerium dioxide (CeO$_2$) with cerium contributing zero nuclear spin and oxygen carrying only 0.04\%\ ($^{17}$O), is a promising potential host for quantum spin $S=\frac{1}{2}$ systems with a theoretically predicted coherence time up to 47 ms \cite{Wolfowicz2021,kanai2022generalized}. Recently, we demonstrated the molecular beam epitaxy of single-crystal Er-doped CeO$_2$ films on Si(111) substrates and the doping dependence on Er$^{3+}$ optical and spin linewidths \cite{Greg2023arxiv}. In this work, we make use of these films with low doping levels (3 parts-per-million (ppm)) to explore the intrinsic optical homogeneous linewidths and electron spin coherence of Er$^{3+}$ in CeO$_2$. Using two pulse photon-echo measurements, we demonstrate that the Er$^{3+}$ ions have long-lived optical states with a narrow homogeneous linewidth of $\sim$ 440 kHz and optical coherence of $\sim$ 0.72 $\mu$s at 3.6 K. Temperature dependent data suggests that the homogeneous linewidth could be $<$ 200 kHz at millikelvin temperatures with optical coherence $>$ 1.6 $\mu$s, indicating the promising potential of Er$^{3+}$ in CeO$_2$ providing a usefully long optical quantum memory. Moreover, the reduced magnetic field noise from a low nuclear spin environment in the CeO$_2$ film enables electron spin polarization with a slow spin-lattice relaxation, thereby enabling access to the electron spin dynamics even at 3.6 K, which is not observable in other, well studied, host materials including Y$_2$SiO$_5$, YVO$_4$, CaWO$_4$ \cite{Raha2020,Welinski2019,Xie2021}. As demonstrated here, the Er$^{3+}$ ions in CeO$_2$ show a spin coherence time $T_2 \sim 0.66$ $\mu s$ at the isolated ion limit with a spin relaxation time $T_1$ $\sim$ 2.5 ms, indicating the potential for millisecond scale spin coherence.

The narrow optical homogeneous linewidth could enable a path towards integration with nanophotonic cavities to drive Er$^{3+}$ optical transitions coherently at even individual ion level to explore time-dependent spectral diffusion \cite{Ourari2023}, critical for entanglement distribution for quantum repeaters. Therefore, the combined narrow optical homogeneous linewidth and long spin relaxation time indicates the potential of such an Er-doped CeO$_2$ platform in providing attributes necessary for efficient optical control of long-lived and coherent spin states for the development of quantum memories.


\section{Results}
\subsection*{Er$^{3+}$ energy structure: crystal field split levels}
The Er$^{3+}$ doped CeO$_2$ sample was epitaxially grown on Si (111)$\pm0.5^{\circ}$ substrate using molecular beam epitaxy (MBE, details in Materials and Methods). A total thickness of 936 nm of single crystal CeO$_2$ with a fluorite structure unit cell was grown and doped with a natural abundance of Er$^{3+}$ isotopes, comprising 77\%\ nuclear-spin-zero even isotopes $^{166}$Er$^{3+}$ and 23\%\ of the odd isotope $^{167}$Er$^{3+}$ with nuclear spin $I=\frac{7}{2}$. The total Er$^{3+}$ concentration is estimated to be 3 ppm based on Er beam flux, and detailed information on growth and structural characterization can be found in prior work \cite{Greg2023arxiv}. 

\begin{figure}[H]
    \centering
    \includegraphics[width=0.95\linewidth]{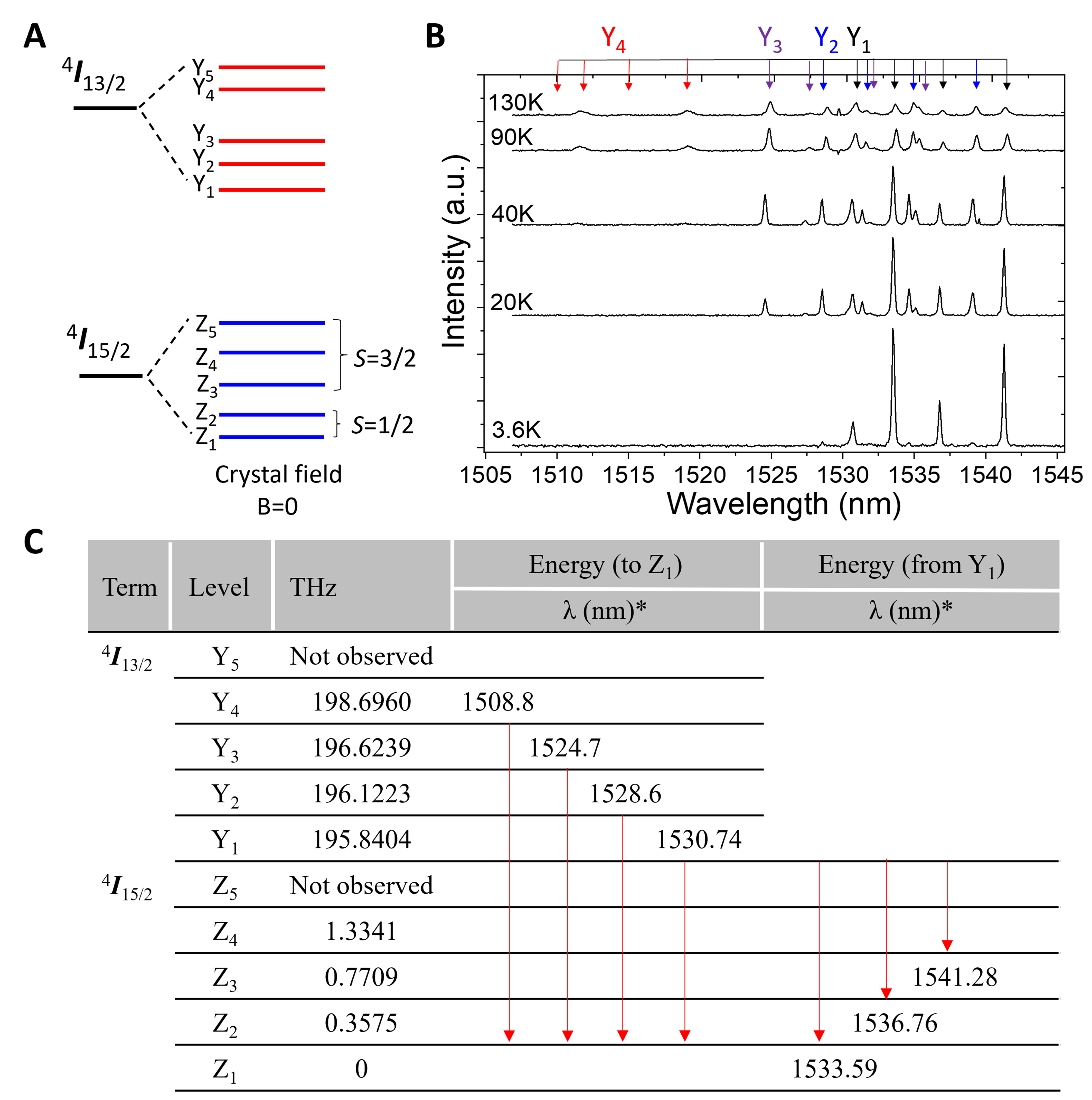}
    \caption{Crystal field split energy levels of the $^4I_{15/2}$ and $^4I_{13/2}$ multiplets of Er$^{3+}$ ions in CeO$_2$. (a) Schematic of the crystal field splitting of the $^4I_{15/2}$ and $^4I_{13/2}$ multiplets, with 5 levels each, labeled as $Z_1$ to $Z_5$ and $Y_1$ to $Y_5$. (b) Temperature dependent PL spectra of Er$^{3+}$ ion emission with Er$^{3+}$ ion excited by a 1473 nm laser with excitation power of 1000 $\mu$W on the sample surface, 5 times the power needed to saturate $Y_1-Z_j$ level transition (details in SI.S2). (c) Table summarizing the crystal field split energy levels.}
    \label{fig:crystal}
\end{figure}

Er$^{3+}$ ions have 11 electrons in the 4f shell that lead to the first two spin-orbit split multiplets as $^4I_{15/2}$ and $^4I_{13/2}$. These multiplets are further split into multiple levels due to the presence of a crystal field. Given the cubic symmetry of the crystal field in CeO$_2$ \cite{Ammerlaan2001,Greg2023arxiv}, the $^4I_{15/2}$ and $^4I_{13/2}$ multiplets split into 5 levels, labeled respectively as $Z_1$ to $Z_5$ and $Y_1$ to $Y_5$, in the order from the lowest to highest energy, as shown in Fig.~\ref{fig:crystal}(a) \cite{Ammerlaan2001}. $Z_1$ and $Z_2$ are two-fold degenerate states with effective spin $S=\frac{1}{2}$ that transform into irreducible representations $\Gamma_6$ and $\Gamma_7$. The higher three Z levels ($Z_3$ to $Z_5$) are four-fold degenerate with effective spin $S=\frac{3}{2}$ that transform into irreducible representation $\Gamma_8$. 

The crystal field split levels of Er$^{3+}$ are probed through temperature and power dependent photoluminescence (PL) measurements with Er$^{3+}$ ions excited by a 1473 nm laser, with photon energy higher than $^4I_{13/2}\rightarrow^4I_{15/2}$ transition, and a spectrometer resolution of 20 GHz resolution (0.16 nm, 84 $\mu$eV). This resolution is sufficient to resolve crystal field split transitions that are typically in the hundreds of GHz to THz range \cite{Liu2004}. At 3.6 K, PL occurs primarily from $Y_1$ to all the Z levels due to the rapid non-radiative relaxation of electrons from higher Y levels to $Y_1$ level. Four emission peaks are observed in Fig.\ref{fig:crystal}(b). These are identified to be $Y_1$ to $Z_1-Z_4$ transitions as marked with black arrows. The lack of emission from the $Z_5$ level may be due to its small transition dipole moment. The higher Y levels are probed by altering the Boltzmann distribution of electrons through increasing the sample temperature from 3.6 K to 150 K shown in Fig. \ref{fig:crystal}(b). Higher Y level transitions are thus identified based on their temperature dependent behavior and their energy separation between each other. With continued increase of temperature, higher Y levels get populated. We observe clearly the $Y_1$ to $Y_4$ levels. The $Y_5$ level transitions may be shorter than 1500 nm and thus not collected in the measurement setup, see the Supplementary Information Section S1 (SI.S1). The intensity of emission from these identified $Y_1$ to $Y_4$ levels also matches the expected behavior from a Boltzmann distribution of electrons at these temperatures (see SI.S2). The table shown in Fig.~\ref{fig:crystal}(c) summarizes the energy structure of the $Z_1$ to $Z_5$ and $Y_1$ to $Y_5$ levels. 

The $Y_1\rightarrow Z_1$ transition is found to be at 1530.74 nm (195.84 THz). The $Y_1$ level is separated from the $Y_2$ level by 1.13 meV (281.9 GHz). Similarly, the $Z_1$ level is separated from the $Z_2$ level by 1.51 meV (357.5 GHz). In this study, we focus on the $Y_1-Z_1$ transition at 1530.74 nm because this transition allows for optical control of the electronic $S=\frac{1}{2}$ spin ground state ($Z_1$ level). As shown later, this transition also has a narrow optical homogeneous linewidth. All studies on optical homogeneous linewidth, on electron spin coherence and relaxation, are carried out at 3.6 K. Given the energy separation between the $Z_1$ and $Z_2$, there is $\approx 0.8$\% electron population of the $Z_2$ level due to Boltzmann statistics. Thus, in all optical measurements resonantly addressing the $Y_1-Z_1$ transition, one can treat the system as an effective two-level system involving only $Z_1$ level and ignore the population of electrons at higher Z levels. However, for spin measurements, the population of the $Z_2$ level becomes more significant and magnifies in the study of spin relaxation dynamics of $Z_1$ level, as will be discussed later. 

\subsection*{Optical coherence of Er$^{3+}$ emission}
\begin{figure}[H]
    \centering
    \includegraphics[width=0.95\linewidth]{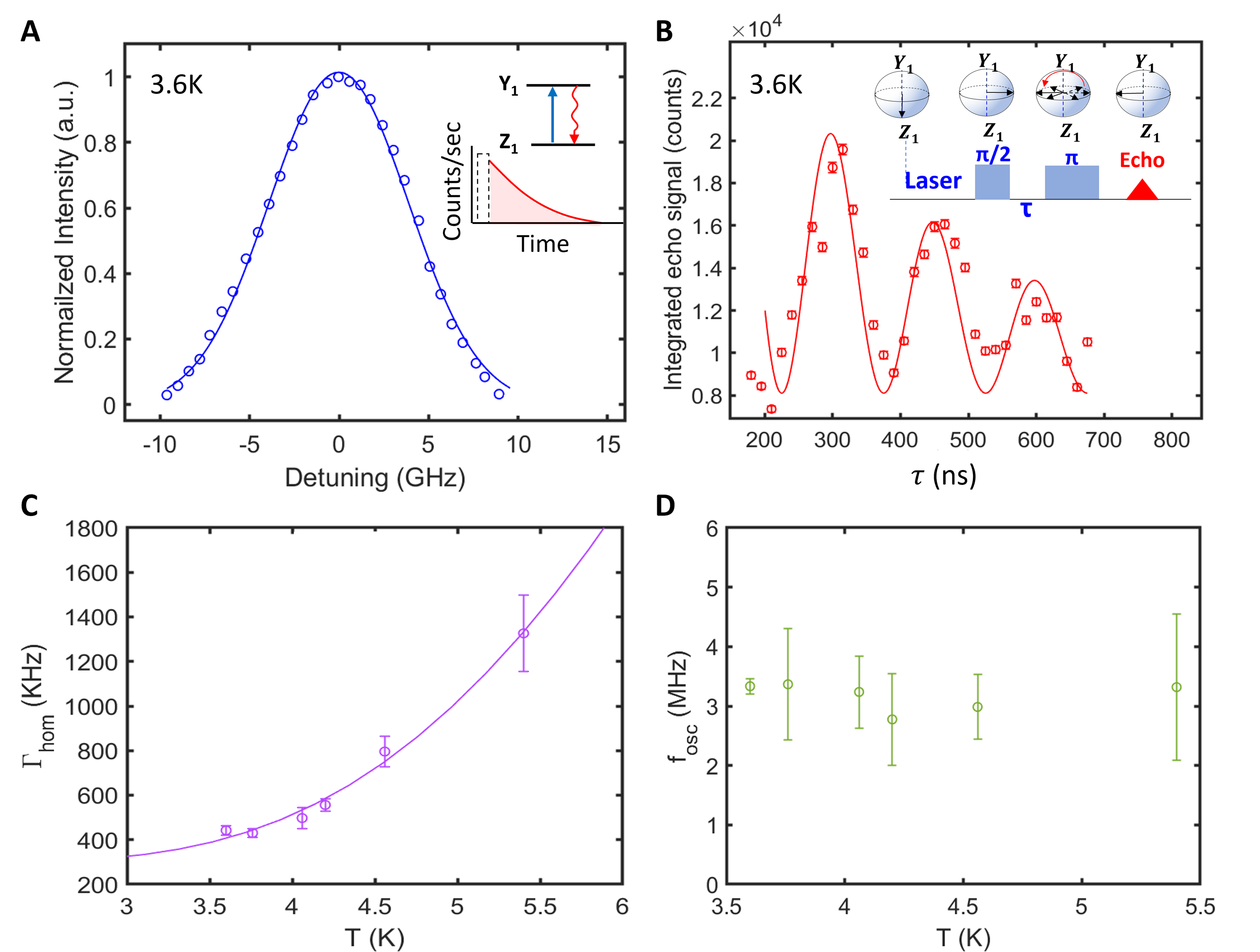}
    \caption{Optical homogeneous linewidth and optical coherence of $Y_1-Z_1$ transition. (a)PLE fine scan of $Y_1-Z_1$ transition with a step size of 0.625 GHz (2.6 $\mu$eV) at 3.6 K with data shown as open circles. The Lorentzian fit to the data is shown as the solid curve, indicating an inhomogeneous linewidth of $\Gamma_{\mathrm{inh}}=9.0\pm0.2$ GHz. The inset shows the schematic of the employed pulse sequence. (b) Two-pulse photon echo (PE) decay for the $Y_1-Z_1$ transition with the pulse sequence shown in the inset with data shown as open circles. The solid line is the fit to the data, indicating optical coherence $T_2=720.0\pm33.1$ ns and homogeneous linewidth $\Gamma_{\mathrm{hom}}=442.1\pm20.3$ kHz. The observed beating pattern is of beating period $T_{\mathrm{osc}}=300.2\pm11.8$ ns ($f_{\mathrm{osc}}=3.33\pm0.23$ MHz). (c) Temperature dependence of $\Gamma_{\mathrm{hom}}$ measured by two-pulse PE with data shown as open circles). The solid line is the fit to the data using Eq.\ref{optT2}. (d) Plot of the extracted $f_{osc}$ (open circles) from two-pulse PE measurement as a function of temperature.}
    \label{fig:opticalT2}
\end{figure}

With the identification of the crystal field split levels of the $^4I_{13/2}\rightarrow^4I_{15/2}$ transition, we focus on the $Y_1\rightarrow Z_1$ transition to probe the inhomogeneous and homogeneous linewidths. The inhomogeneous linewidth is probed using photoluminescence excitation spectroscopy (PLE), the measured spectrum is shown in  Fig.\ref{fig:opticalT2}(a). It is obtained using the optical pulse sequence schematically shown in the inset of Fig.\ref{fig:opticalT2}(a) (details see Materials and Methods) while scanning the laser frequency across the $Y_1-Z_1$ transition with a step size of 0.625 GHz. The Lorentzian fit to the data indicates an inhomogeneous linewidth of $\Gamma_{\mathrm{inh}}=9.0\pm0.2$ GHz (0.07 nm or 37 $\mu$eV), comparable to other MBE grown Er$^{3+}$ doped in other rare-earth oxide films such as Y$_2$O$_3$ and TiO$_2$\cite{Singh2020, Dibos2022}. Compared with Er$^{3+}$ ions in other bulk low nuclear spin bath host materials, such as YSO (Y$_2$SiO$_5$), Y$_2$O$_3$, and CaW$O_4$, the observed linewidth is around a factor of ten higher \cite{Sun2002,Stevenson2022}. This is likely due to the relative high density of threading dislocations and unintended defects in the epitaxial CeO$_2$ film on Si originating from the 0.5\%\ lattice mismatch strain \cite{Greg2023arxiv}. The observed signal is dominantly from $^{166}$Er ions. The emission from the 23\%\ of $^{167}$Er is buried under the observed broad inhomogeneous peak. Therefore, we are unable to resolve the hyperfine splitting from $^{167}$Er.

Besides inhomogeneous linewidth, the homogeneous linewidth is another important figure-of-merit for an optical transition. One can extract the homogeneous linewidth through the measurement of optical coherence ($T_2$) where $\Gamma_{\mathrm{hom}}=\frac{1}{\pi T_2}$ \cite{Abella1966}, and we do so via two-pulse photon-echo (PE) measurement to probe the homogeneous linewidth. Fig.\ref{fig:opticalT2}(b) shows the measured integrated echo intensity as a function of $\tau$ at 3.6 K using the pulse sequence schematically shown in the inset (details in Materials and Methods and SI.S3). The data show a single exponential decay envelope of the photon echo amplitude modulated with an oscillating beat pattern. The beating pattern indicates that we are coherently addressing of a superposition of two transitions in a three-level system with the energy separation of two of the levels being within the bandwidth of the optical pulse. The red line is a fit to the data considering a single exponential decay with an added frequency of oscillation $f=1/T_{\mathrm{osc}}$. The data indicate an optical coherence $T_2=720.0\pm33.1$ ns with homogeneous linewidth $\Gamma_{\mathrm{hom}}=\frac{1}{\pi T_2}=442.1\pm20.3$ kHz and a beating period of $T_{\mathrm{osc}}=300.2\pm11.8$ ns ($f_{\mathrm{osc}}=3.33\pm0.23$ MHz). The observed beating frequency is within the bandwidth of the optical pulse and also found to be consistent with the Zeeman splitting of the $Z_1$ level due to earth's magnetic field at around 0.35 G. This suggests that the beating might be from the effect of earth's magnetic field lifting the degeneracy of $Z_1$ level. 

The observed homogeneous linewidth, $\Gamma_{\mathrm{hom}}=442.1\pm20.3$ kHz, at 3.6 K without any externally applied magnetic field is orders of magnitude higher compared to the lifetime-limited $\Gamma_{\mathrm{hom}}\sim94$ Hz given the $\sim$3.4 ms radiative lifetime \cite{Greg2023arxiv} of the $Y_1-Z_1$ transition. To probe the dephasing processes, temperature dependent measurements of $\Gamma_{\mathrm{hom}}$ is carried out using two pulse PE to gain insights into the dephasing mechanisms occurring in the material, with temperatures ranging from 3.6 K to 5.5 K. Fig.\ref{fig:opticalT2}(c) shows the extracted $\Gamma_{\mathrm{hom}}$ as a function of temperature. The beat frequency extracted from all temperatures is shown in Fig. Fig.\ref{fig:opticalT2}(d). The beat frequency is independent of temperature which is consistent with its origin being from the Zeeman splitting of the $Z_1$ level induced by the earth's magnetic field. At these measured temperatures, two phonon processes contribute to dephasing: (a) coupling to two-level systems (TLS) \cite{Anderson1972,Huber1984,Flinn1994} (b) Orbach process phonon effects \cite{Flurin2003,Sun2012} with a homogeneous linewidth $\Gamma_{\mathrm{hom}}(T)$ of the following form:

\begin{equation}
    \Gamma_{hom} (T) =\Gamma_0+\alpha_{\mathrm{TLS}}\cdot T+\alpha_{\mathrm{phonon}}\cdot exp(\frac{-\Delta E}{K_BT})
      \label{optT2}
\end{equation}

\noindent where $\Gamma_0$ is the linewidth at 0 K, $\alpha_{TLS}$ is the coefficient for coupling to TLS, $K_B$ is the Boltzmann constant. 

In the probed temperature range, the increase in $\Gamma_{\mathrm{hom}}$ is dominated by Orbach relaxation. The solid line is the fit to the data using Eq.\ref{optT2} with $\Delta E=2.05$ meV. The extracted $\Delta E$ is consistent with the energy separation between the $Z_1$ to $Z_2$ level obtained from PL measurements. Of the total linewidth broadening, $\approx150$ kHz is due an Orbach process at 3.6 K with the remaining 300 kHz of broadening coming from the combined contribution of $\Gamma_0$ and direct phonon coupling, $\alpha_{\mathrm{phonon}}\cdot T$. The $\alpha_{\mathrm{phonon}}$ is typically in the range of a few to tens of kHz/K \cite{Serrano2022, Fukumori2020,Flurin2003} for rare-earth ions in oxides. One can thus deduce that $\Gamma_0$ is most likely $\leq$ 200 kHz. This suggests that the dominant dephasing process might be from spectral diffusion due to ion-ion dipolar interactions given the short ensemble average Er-Er separation in the sample ($\sim$14 nm, estimated from the Er concentration) or a fluctuating field induced by background charge and defects as well as strain in the film. It is worth noting that the sample studied here is grown without any optimization. One can further improve on the homogeneous linewidth by optimizing growth to reduce strain and minimize defects. There is also the path of reducing the concentration of Er$^{3+}$ to minimize ion-ion dipolar interaction induced spectral diffusion. Besides this, one can also improve the homogeneous linewidth by applying moderate magnetic field to reduce the coupling of TLS to the dipole moment of Er$^{3+}$ \cite{Sun2002,Flurin2003}. This can lead to lower tunneling rate, thus reducing the magnetic noise caused by TLS. The field can also freeze Er spin flip-flop processes to reduce fluctuating magnetic field induced spectral diffusion and thus extend optical coherence. 

The obtained homogeneous linewidth $\Gamma_{\mathrm{hom}}=442.1\pm20.3$ kHz at 3.6 K suggests that a viable path exists for further engineering the light-matter interaction via integrating Er ions in a cavity with a suitable quality factor to enhance the radiative transition rate through Purcell effect close to its optical coherence limit \cite{Dibos2018, Zhong2018}. Further improvement on the homogeneous linewidth through growth optimization and reduction of spectral diffusion and phonon mediated dephasing at millikelvin temperature with external magnetic field can aid in reaching sub-kHz homogeneous linewidths towards radiative-lifetime limited coherent photon generation, which are needed coherent optical control of the Er spin states.    

\subsection*{Er$^{3+}$ electron spin coherence}
\begin{figure}[H]
    \centering
    \includegraphics[width=0.95\linewidth]{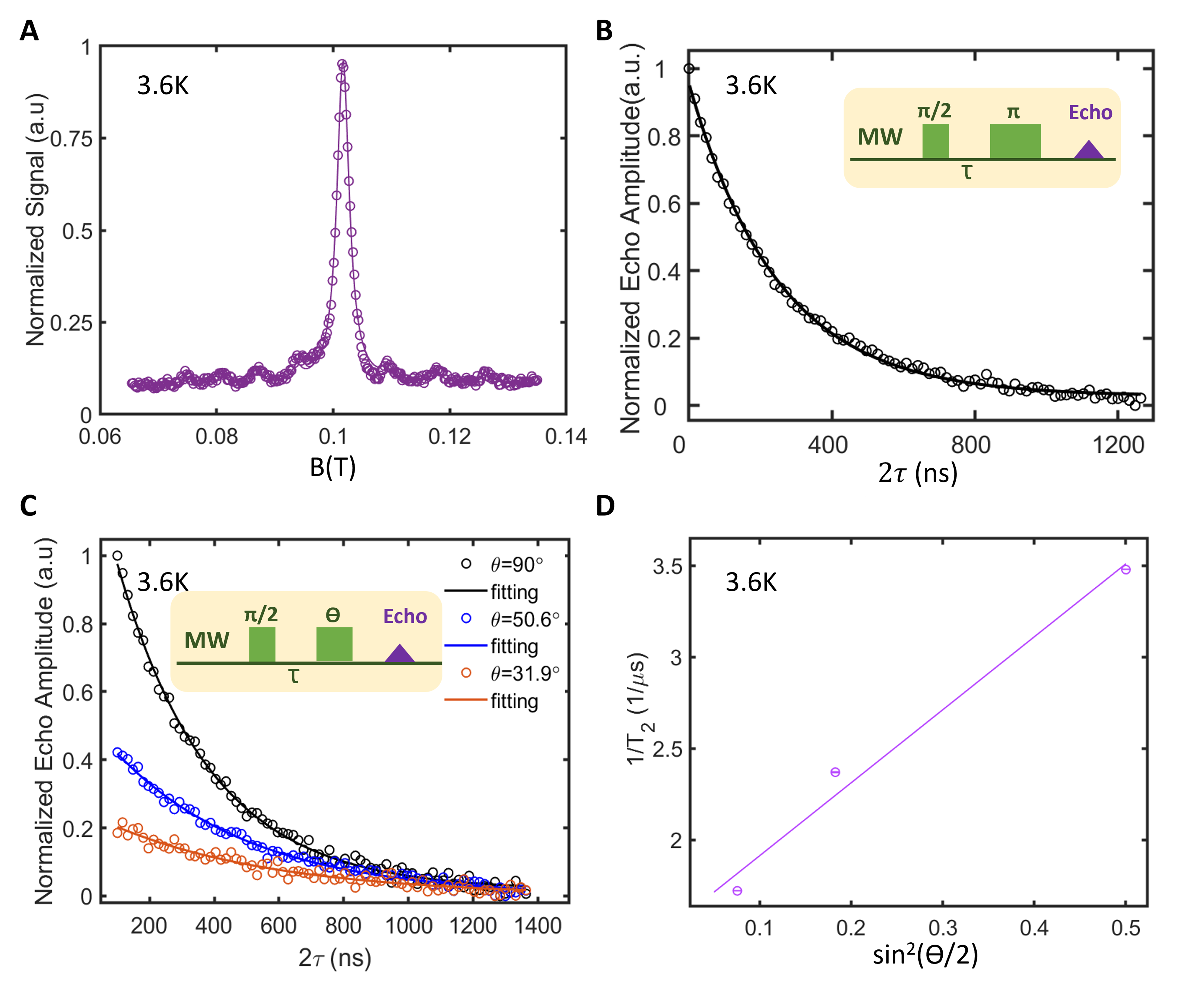}
    \caption{Electron spin coherence at 3.6 K probed by pulsed EPR with microwave drive at 9.7 GHz (a) Resultant EPR spectrum from two-pulse echo as the magnetic field (B) is swept. Data is taken using the Hahn-echo sequence schematically shown in panel (b) with fixed delay time ($\tau=100$ ns) between the $\pi/2$ pulse and $\pi$ pulse. (b) Spin echo measurement using two-pulse Hahn-echo sequence schematically shown in the inset. Data (open circles) are taken as a function of time delay $\tau$ between the 12ns $\pi/2$ pulse and 24ns $\pi$ pulse. The solid black line is a single exponential fit revealing the spin coherence time $T_2=0.249\pm0.035$ $\mu s$ (c) Generalized Hahn echo measurements on Er$^{3+}$ electron spins taken with three different flip angles $\theta$ of the second rotation pulse. Solid lines are the single exponential fits to the data. (d) Plot of the inverse of spin coherence time $T_2$, extracted from the fits to the data in panel (c) as a function of averaged inversion pulse fidelity $<{\mathrm{sin}}^2(\theta/2)>$. Linear fit to the data (solid line) yields the spin coherence at single isolated ion limits to be $T_2=0.660$ $\mu s$ and a spin concentration of $5.66\pm0.25$ ppm in the sample.}
    \label{fig:EPR}
\end{figure}

The optical coherence study discussed earlier indicates that Er$^{3+}$ ions in CeO$_2$ show a narrow optical homogeneous linewidth with promising potential of providing up to several microseconds of optical coherence at millikelvin temperatures. For the application and use of Er$^{3+}$ as the spin-photon interface needed for quantum memory, the Er$^{3+}$ electron spin coherence is the other important figure-of-merit. Next, we move on to study the Er$^{3+}$ electron spin behavior. We use a X-band (9.7 GHz) pulsed electron paramagnetic resonance (EPR) spectrometer to study spin coherence and relaxation. Fig.~\ref{fig:EPR} (a) shows the measured spin echo response as a function of the static magnetic field. The data are taken with the $\tau$ delay between the $\pi/2$ (12 ns) and $\pi$ (24 ns) pulses at 100 ns. The resultant spectrum shows the expected resonance from nuclear-spin-zero even isotopes $^{166}$Er$^{3+}$ (primary peak) along with the hyperfine levels of 23\%\ of the $^{167}$Er$^{3+}$ with nuclear spin $I=\frac{7}{2}$ (smaller secondary peaks), consistent with results in prior work \cite{Greg2023arxiv}.  Seven of the eight hyperfine peaks are spectrally clearly resolved with one of the hyperfine resonance hiding under the primary resonance peak at $B_0=0.102$ T. The measured spectrum is fit using Zeeman and hyperfine terms while also accounting for the second-order perturbation effects from the large nuclear spin of $^{167}$Er$^{3+}$. The extracted g value is $g=6.828\pm0.005$, consistent with the CW EPR work reported in prior work \cite{Greg2023arxiv}. The g value matches theoretical estimates and reported values for Er$^{3+}$ in CeO$_2$ nanocrystals (\cite{Ammerlaan2001,Rakhmatullin2014}). The obtained linewidth of the resonance of Er$^{3+}$ spin is $2.57\pm0.03$ mT($244.9\pm2.9$ MHz), consistent with CW results\cite{Greg2023arxiv}. This result confirms the expected EPR resonance of Er$^{3+}$ ion in a crystal with cubic crystal field symmetry. 

The Er$^{3+}$ spin coherence time, $T_2$, is probed via Hahn-echo measurement. The magnetic field is tuned to its resonance at $B=0.102$ T, the resonance of the primary peak shown in Fig.~\ref{fig:EPR} (a) under the applied 9.7 GHz microwave frequency. The echo signal is predominantly from the $^{166}$Er electrons with only 3.6\%\ of the signal from the $^{167}$Er electrons. Fig.~\ref{fig:EPR} (b) shows the measured spin echo signal collected as a function of the delay ($\tau$) between the $\pi/2$- and the $\pi$-pulse at 3.6 K.  From the fit of $I\propto {\mathrm{exp}}(-T_2/2\tau)$ (black curve in Fig.~\ref{fig:EPR} (b)), one obtains $T_2=0.249\pm0.035$ $\mu s$. The spin $T_2$ is typically limited by the phonon induced dephasing at such elevated temperatures and Er-Er spin dipolar interactions. The Er-Er dipolar spin interaction shifts the spin resonance. These shifts fluctuate and spectral diffusion occurs because of the random spin orientation resulting from spin-lattice interactions or spin diffusion. The dipolar interaction between spins magnifies itself in the spin echo decay through so-called instantaneous diffusion \cite{Salikhov1981,Eichhorn2019}. The spin coherence can be written as $1/T_2=1/T_{\mathrm{2,INST}} + 1/T_{\mathrm{2,bath}}$ where $T_{\mathrm{2,INST}}$ represents the contribution from instantaneous diffusion. To understand the dominant dephasing mechanics, we carry out instantaneous diffusion measurements \cite{Salikhov1981} to probe and decouple the Er-Er spin dipolar interactions. A generalized Hahn echo  sequence ($\pi/2 -\tau- \theta$) is performed  on  Er  spins  while  the  angle,  and  hence the fidelity, of the second inversion pulse is varied \cite{Salikhov1981,Eichhorn2019}. The second pulse inhibits the decoupling of the probed spins’ mutual dipolar interactions, resulting in decoherence through instantaneous diffusion. The echo signal (SE) is thus proportional to the exponential of the averaged inversion pulse fidelity $<{\mathrm{sin}}^2(\theta/2)>$ and is given by \cite{Salikhov1981}:

\begin{equation}
    SE(\tau) \propto exp(\frac{8\pi^2}{9\sqrt(3)}\frac{g^2\beta^2}{\hbar}N{\mathrm{sin}}^2(\theta/2)\tau)
      \label{EqSE}
\end{equation}

\noindent Thus, $T_{\mathrm{2,INST}}$ is proportional $<{\mathrm{sin}}^2(\theta/2)>$ and one have the following equation for $T_2$ where N is the total number of spins per $m_3$ and $\beta$ is the Bohr magnon, 

\begin{equation}
    1/T_2=1/T_{\mathrm{2,INST}}+1/T_{\mathrm{2,bath}}=\frac{8\pi^2}{9\sqrt(3)}\frac{g^2\beta^2}{\hbar}N{\mathrm{sin}}^2(\theta/2)+1/T_{\mathrm{2,bath}}
      \label{EqT2}
\end{equation}
 
In the instantaneous diffusion measurements, the angle of the second rotation pulse $\theta$ is varied by tuning the power of the microwave pulse (supplementary Section S4), while keeping the pulse length unchanged so that one rotates the same ensemble of spin within the second pulse. Fig.~\ref{fig:EPR} (c) shows the measured echo intensity as a function of $\tau$ with three different rotation angles $\theta$. A reduction of rotation angle $\theta$ reduces spin flips induced by the microwave pulse, hence, reduces instantaneous diffusion. Er spin $T_2$ increased from $0.25\mu s$ to $0.58 \mu s$. The inverse of the extracted $T_2$ obtained through the single exponential fit is shown in Fig.~\ref{fig:EPR} (d). Following Eq.~\ref{EqT2}, the slope of the linear fit to the data in Fig.~\ref{fig:EPR} (d) yields the density of probed Er spins to be $1.66\pm0.08*10^{22}/m^3$, $0.68\pm0.03$ ppm. To estimate the overall concentration of Er, we need to take into account the fraction of probed Er out of the entire ensemble. The linewidth of the $Er^{3+}$ spin resonance is $2.57\pm0.03$ mT ($244.9\pm2.9z$) MHz, around 8.3 times larger than the bandwidth of the $\theta$ rotation pulse. This indicates that only 12\%\ of the spins within the inhomogeneous distribution are probed. Therefore, the estimated total concentration of the Er spin is $5.66\pm0.25$ ppm, within a factor of 2 of the Er concentration estimated from Er flux used during MBE growth. The intercept of the linear fit provides an estimate on the spin coherence at the single isolated ion limit with $T_2=T_{\mathrm{2,bath}}=0.660\pm0.004$ $\mu s$. Thus, the measured $T_2$ in Fig.~\ref{fig:EPR} (b) is largely limited by the Er-Er spin dipolar interaction induced instantaneous diffusion and could thus be improved by a reduction of Er concentration. With the generalized echo sequence reducing instantaneous diffusion, the spin homogeneity $\Gamma_{\mathrm{h}}=\frac{1}{\pi T_2}$ contributed by the bath is $484.8\pm20.6$ kHz. The deduced spin coherence $T_{\mathrm{2,bath}}$ at the single isolated ion limit is probably limited by the phonon induced dephasing and spectral diffusion induced by interaction with other defects in the film. Further work on studying spin $T_2$ at lower temperature to further probe the nature of dephasing dynamics is underway. 

\subsection*{Er$^{3+}$ electron spin relaxation}
\begin{figure}[H]
    \centering
    \includegraphics[width=0.95\linewidth]{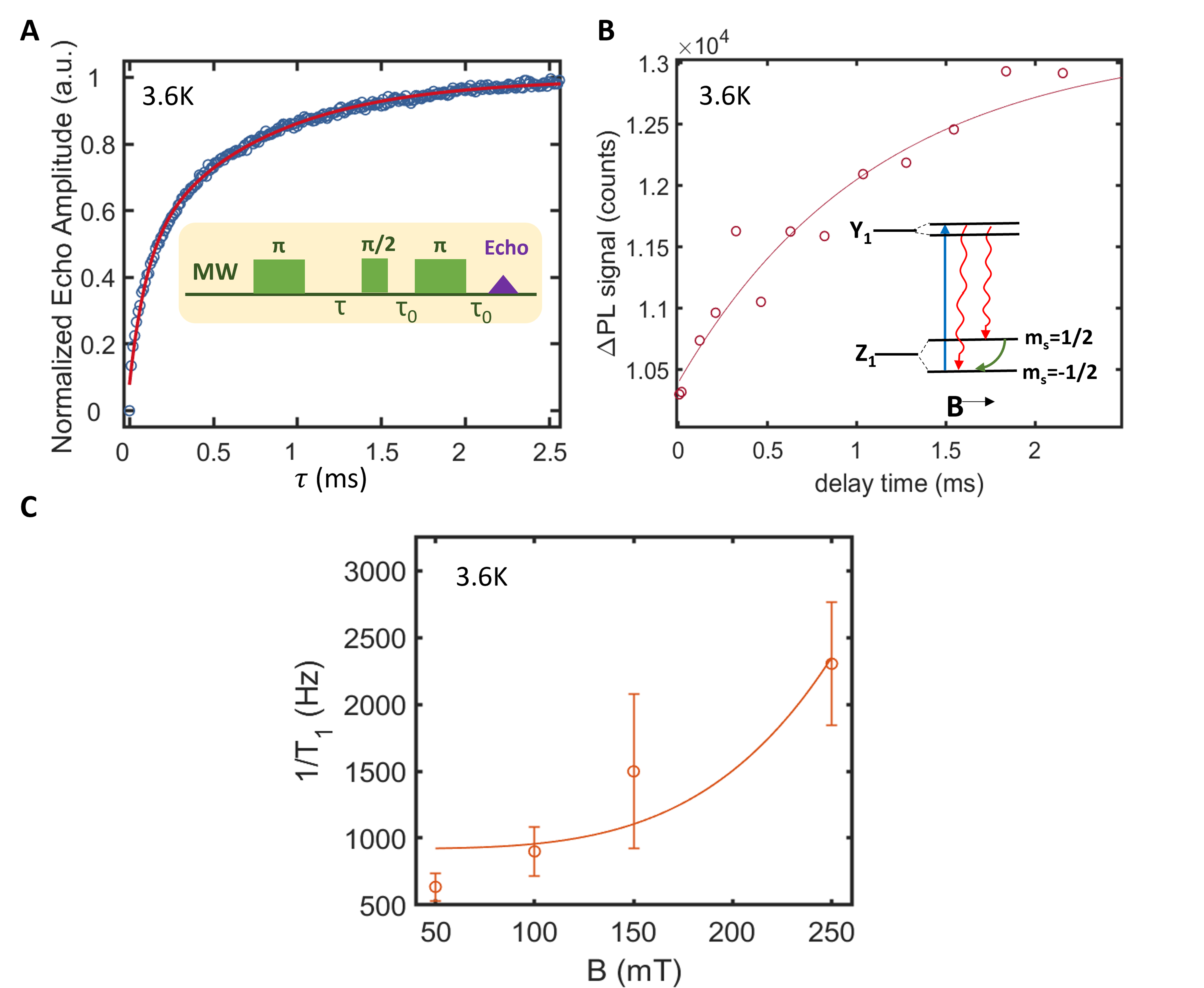}
    \caption{Electron spin relaxation dynamics at 3.6 K (a) Pulsed EPR based spin echo measurement with a three-pulse population inversion sequence (see Methods and Materials) shown in the inset. The measured data are shown as open circles and the solid line is a double exponential fit revealing two spin relaxation paths with a short spin relaxation $T_1=0.11\pm0.01$ ms and a long relaxation $T_10.83\pm0.04$ ms. (b) Optical measurement of $Z_1$ level spin relaxation using two optical pulses using 'pump and probe' scheme. The panel shows the measured PL signal difference ($\Delta {\mathrm{PL}}$, open circles) with and without the second probe optical pulse as a function of time delay $\tau$ between pulses. Data are taken with laser resonant to the spin-down $Z-1$ to spin-up $Y_1$ level transition (inset). A 100 $\mu s$ pulse is applied first followed with a second 100 $\mu s$ pulse separated by time $\tau$. PL signal is collected within a 4 ms collection window with 15,000 iterations of measurements. Details on the pulse sequence is in Supplementary Information Section S5. (c) Optically measured spin relaxation time $T_1$ as a function of the applied magnetic field. The plotted $T_1$ is the fitted value with 95\% confidence extracted from each measured $\Delta {\mathrm{PL}}$ data in the SI.S5. The data show an increase of 0.437 ms to 1.575 ms by reducing the field strength from 250 mT to 50 mT.}
    \label{fig:OptSpinT1}
\end{figure}

The limit on spin coherence is set by the spin relaxation. One can probe the spin relaxation mechanism to obtain an upper limit on the spin $T_2$. The spin relaxation time is probed by first applying a $\pi$-pulse to inverse the population of spin-up and spin-down electron states, and then reading out the relaxation of spin-up to spin-down using the two-pulse Haho-echo sequence (schematically shown in Fig.\ref{fig:OptSpinT1}(a), details see Methods and Materials). By varying the delay time, $\tau$, between the inversion pulse and the $\pi/2$-pulse, one can then map out the spin relaxation dynamics. Fig.~\ref{fig:OptSpinT1} (a) shows the measured spin echo signal as a function of $\tau$ measured at 3.6 K. The data indicates the presence of two spin relaxation channels with a short spin relaxation $T_1=0.11\pm0.01$ ms and a long relaxation $T_1=0.83\pm0.04$ ms. The two observed decay processes might come from the electron depopulation of the $Z_1$ spin level to its nearby $Z_2$ level, resulting in a sampling of electron population between three active states. EPR is an inductive detection method that is sensitive to the population of ground states. At 3.6 K, there is thermal population of both $Z_1$ and $Z_2$ levels given the small energy separation of 1.51 meV (357.5 GHz). The added Zeeman splitting further aids in reducing the energy barrier between $Z_1$ spin-up and $Z_2$ spin-down levels. Possible depopulation of $Z_1$ spin-up level to $Z_2$ spin level, mediated by phonon processes, can thus be detected by pulsed EPR.  

To further probe the origin of the observed double exponential decay dynamics, we carry out optical measurements of spin relaxation $T_1$ of the $Z_1$ level at 3.6 K. We apply a 100 mT magnetic field parallel to the $\langle1 -1 0\rangle$ direction, with orientation chosen to be the same as that used in the pulsed EPR measurements. A Zeeman splitting of 9.46 GHz between the spin-up and spin-down state of the $Z_1$ level is induced, estimated based on the effective g value extracted from data shown in Fig. \ref{fig:EPR}(a). Given that the spin $T_1$ is shorter compared to the optical lifetime of the $Y_1-Z_1$ transition, the optical measurement of spin $T_1$ cannot be done using typical spectral hole-burning method \cite{Car2019,Hastings-Simon2008}, where one can fully polarizes the spins through the cumulative optical excitation processes. Here we use a two pulse optical based 'pump-probe' scheme to probe the spin relaxation from spin up to spin down state that magnifies in population inversion recovery of the spin down states. We first apply a short 100 $\mu s$ optical pulse resonant to the transition between $Z_1$ spin down state and the $Y_1$ spin up state (inset of Fig.\ref{fig:OptSpinT1} to drive the electrons occupying spin-down states to the excited state, creating an initial state occupation where the population of the $Z_1$ spin-up state is higher than that of spin-down. A second pulse of 100 $\mu s$ is applied after a delay, $\tau$, to probe the recovery of the spin-down state occupation due to spin relaxation. A reference measurement without the second excitation pulse is taken to sample the photon emission into collection window from optical decay of the $Y_1$ level to both spin states after the first optical pulse as the background signal for subtraction  (details in SI.S5).

Figure \ref{fig:OptSpinT1}(b) shows the measured difference of the PL signal, $\Delta {\mathrm{PL}}$, collected during the collection window with and without the second optical pulse as a function of $\tau$ between the two optical pulses. The spin recovery from spin-up to spin-down through spin relaxation is evidenced by the increasing $\Delta {\mathrm{PL}}$ with increasing $\tau$. The data shows a single exponential decay (fitting, solid line) indicating a spin relaxation time for the Zeeman split $Z_1$ spin-up to spin-down state of $T_1=1.106\pm0.256$ ms. The measured $T_1$ value is consistent with the long $T_1$ resolved in the pulsed EPR shown in Fig.\ref{fig:OptSpinT1}(a). The observed single exponential decay in optical measurement of spin relaxation of $Z_1$ level also suggests that the short $T_1$ of 0.11 ms observed in pulsed EPR measurement is most likely coming from the phonon mediated depopulation of electrons from $Z_1$ spin up level to $Z_2$ level. 

At this temperature, the spin relaxation time $T_1$ of the ground state is limited by phonon-mediated processes, including direct, Raman and Orbach processes \cite{AbragamEPR1986}. One can further extend the relaxation time by tuning the magnetic field to control the direct coupling process, which can be suppressed with lower magnetic field by reducing the number of phonon modes that can couple to the Zeeman split states as \cite{Lutz2016}: 
\begin{equation}
    T_1^{-1} = A_{\mathrm{o}}(g^4){\mathrm{sech}}^2(\frac{g\mu_B B}{2k_B T})+A_{\mathrm{d}}(\frac{g\mu_B B}{h})^5{\mathrm{coth}}(\frac{g\mu_B B}{2k_B T})+R_{\mathrm{o}}
      \label{EqT1}
\end{equation}
where $g\mu_B B$ is the energy difference between the two $Z_1$ spin sub-levels under magnetic field $B$, $k_B$ is Boltzmann's constant and $h$ is Planck's constant. Figure \ref{fig:OptSpinT1}(c) shows the measured spin $T_1$ time as a function of the magnitude of the applied magnetic field. We measure $T_1$ with B-field at 50 mT, 100 mT, 150 mT and 250 mT. The $T_1$ values are extracted from measured $\Delta {\mathrm{PL}}$ data (see SI.S5) using the same two optical pulse sequence. We observe an extension of $T_1$ from $T_1=1.106\pm0.256$ ms to $T_1=1.575\pm0.256$ ms when reducing the field strength from 100 mT to 50 mT, and similarly a reduction of $T_1$ to $0.4345\pm0.087$ ms at an elevated field of 250 mT. The solid line shown is a fit to the $T_1$ data using Eq.\ref{EqT1} with the Raman and Orbach processes treated as a constant in the fitting. The fitting suggests that one can further extend the spin $T_1$ to 2.5 ms at 3.6 K. The observed spin $T_2$ in the single-ion limit is 0.66 $\mu s$, much less than the observed spin $T_1$, possibly due to magnetic and electric noise from Er spin flip induced spectral diffusion and defects present in the film. One could further extend both spin $T_1$ and $T_2$ by freezing out the spin-flip induced dephasing at moderate fields and freezing out higher order phonon effects at millikelvin temperatures.


\section{Conclusion}

Our work on Er-doped CeO$_2$ hightlights the potential of this material system as a robust optical quantum memory platform owing to its narrow linewidth and long-lived optically addressable electron spin, enabled by the low nuclear spin host environment. The observed homogeneous linewidth of 440 kHz for the $Y_1-Z_1$ transition and electron spin relaxation time of 2.5 ms at 3.6 K indicate the feasibility of using collective electron spin relaxation as a local quantum memory for quantum repeaters. The narrow homogeneous linewidth of 440 kHz also demonstrates the potential for integrating Er$^{3+}$ with nanophotonic cavities to achieve Purcell enhancement and near Fourier transformation-limited single-photon emission. This would allow for coherently driven optical transitions at a desired rate to address individual ions \cite{Dibos2018,Kindem2020} and examine the time-dependent spectral diffusion of individual Er$^{3+}$ ions in the host \cite{Ourari2023}, a critical step towards entanglement distribution needed for quantum repeaters. The significant reduction in the concentration of nuclear magnetic moments in CeO$_2$ compared to that of other hosts, such as Y$_2$SiO$_5$ and YAG, could open a path towards not only long-lived coherent Er$^{3+}$ electron spin states, but also long-lived nuclear spins in isotopically enriched $^{167}$Er to enable long storage times on the scale of seconds, using both collective relaxation modes of nuclear spin ensembles \cite{Rancic2018} and individual
nuclear spin states \cite{Jiang2009,Lee2013}. 

The Er$^{3+}$ spin ensemble coherence value reported here is largely limited by the ion-ion dipolar interaction. As indicated by the instantaneous diffusion measurements, in the single-ion limit, the Er$^{3+}$ spin coherence $T_2$ is around 0.66 $\mu s$. One can be further improve spin coherence by lowering the temperature below the explored 3.6 K in this work to millikelvin temperature. One can also further improve coherence by using  higher magnetic fields to freeze spin-flip induced dephasing. The MBE growth of Er-doped oxides also enables the control of Er doping levels and optimization of material quality in minimizing defects and dislocations in the film to reach high quality single crystal CeO$_2$ thereby reducing spectral diffusion and improving on both optical and spin properties. It can also enable growth of CeO$_2$ thin films with controlled delta doping of Er to create structures compatible with integration with nanophotonic cavities, either within the oxide or hybrid structures integrated with other dielectric materials. The sample studied here is grown without growth optimization yet already demonstrates appreciable spin relaxation ( $\sim$2.5 ms) and narrow optical homogeneous linewidths (440 kHz). Continued growth optimization employing slower growth rates\textemdash with a lower oxygen pressure \cite{Nishikawa2002} to suppress formation of dislocations and further reduce unintended defect concentrations in the film\textemdash has the potential to significantly reach narrower homogeneous linewidths and longer spin coherence and relaxation. Thus, Er$^{3+}$ in CeO$_2$, an oxide host with a very low nuclear spin environment, could emerge as a versatile platform for highly coherent light–matter quantum interfaces for developing quantum communication applications.


\section{Materials and Methods}
\subsection*{Sample and growth}
The Er-doped CeO$_2$ film studied here is grown on silicon (111)$\pm$0.5$^{\circ}$ using molecular beam epitaxy. Single crystal CeO$_2$ is grown at a sample temperature of $670^{\circ}C$ with a growth rate of 312.1 nm/hr under oxygen pressure of 4.9e-6 torr and Ce/$O_2$ flux ratio $\sim20$. The grown CeO$_2$ layer is 936.3 nm thick with Er-doped through the entire grown layer with an estimated Er concentration of 3ppm, based on the Erbium flux delivered during growth \cite{Greg2023arxiv}. Details on growth condition and structural characterization of as-grown films can be found in Ref.\cite{Greg2023arxiv}. The as-grown sample is directly used for all the measurements shown here without any post-growth processing. 

\subsection*{Optical pulse sequences for optical measurements}
All optical data shown in the main text are collected at 3.6 K with Er emission collected using time gating method. For PLE measurements (Fig. \ref{fig:opticalT2}(a)), 1.5 ms long optical pulse with wavelength tuned across $Y_1-Z_1$ transition is used to excite the Er ions. A collection window of 7 ms after the excitation pulse is used to collect the emission from Er ions. The collection window is chosen based on the optical lifetime of the $Y_1-Z_1$ transition (3.4 ms, Ref. \cite{Greg2023arxiv}) to enable needed signal-to-noise. For photon echo measurements (Fig.\ref{fig:opticalT2}(b) and (c)), 10 ns $\pi/2$-pulse followed with a 20 ns $\pi$-pulse after a delay $\tau$ is used. The laser is set to be resonant with the $Y_1-Z_1$ transition with laser power tuned to reach needed $\pi/2$ and $\pi$ pulse area (see SI.S3). The shortest possible pulse enabled by our instrumentation is used to minimize dephasing during the excitation process. Data are collected with $\tau$ ranging from 180 ns to 700 ns. Details on instrumentation for all optical measurements are captured in SI.S1 and SI.S3.    

\subsection*{Pulsed EPR measurements}
X-Band (9.7 GHz) EPR experiments are taken using an ELEXSYS E580 spectrometer (Bruker Biospin, Ettlingen, Germany) that is equipped with a dielectric ring resonator (Bruker ER 4118X-MD5). The Er-doped CeO$_2$ film on Si samples are diced to a size of 4 mm x 2.5 cm and mounted into a quartz tube suspended in the center of the dielectric ring resonator contained in a flow cryostat (Oxford Instruments CF935) with pumped liquid helium. The data shown in the manuscript are obtained at 3.6K with temperature controlled by an ITX temperature controller (Oxford Instruments). 
For spin echo field sweep (Fig.\ref{fig:EPR}(a)), two-pulse Hahn echo sequence is applied with a 12 ns $\pi/2$-pulse followed by a 24 ns $\pi$-pulse with a fixed delay $\tau$ of 100 ns. The pulse length is chosen for the shorted achievable length to cover large size of Er spin ensemble. For spin coherence measurement (Fig.\ref{fig:EPR}(b)), the same Hahn echo sequence is used with a varying delay $\tau$ ranging from 100 ns to 1300 ns. For spin relaxation measurement (Fig.\ref{fig:OptSpinT1}(a)), three-pulse population inversion sequence is used where a 24 ns $\pi$-pulse is followed by a two-pulse Hahn echo sequence with varying time delay $\tau$. The two-pulse Hahn echo sequence used here is composed of a 12 ns $\pi/2$-pulse followed by a 24 ns $\pi$-pulse with a fixed delay $\tau$ of 100 ns.

\section{Acknowledgements}
The authors would like to thank Dr. Jonathan Marcks and Dr. Yeghishe Tsaturyan for helpful discussions. This work was primarily funded (J.Z., M.T.S., F. J. H., D. D. A.) by the  U.S. Department of Energy, Office of Science, Basic Energy Sciences, Materials Sciences and Engineering Division, including support for optical and spin characterization studies. The sample growth (I.M., G.D.G., S.G.) along with additional support for cryo-optical measurements (G.D.G., A.M.D.) was funded by Q-NEXT, a U.S. Department of Energy Office of Science National Quantum Information Science Research Centers under Award Number DE-FOA-0002253. The EPR work in the Chemical Sciences and Engineering Division was supported by the U.S. Department of Energy, Office of Science, Office of Basic Energy Sciences, Division of Chemical Sciences, Geosciences, and Biosciences, through Argonne National Laboratory under Contract No. DE-AC02-06CH11357.

\subsection*{Funding:}
U.S. Department of Energy, Office of Science, Basic Energy Sciences, Materials Sciences and Engineering Division
U.S. Department of Energy, Office of Science, National Quantum Information Science Research Centers
U.S. Department of Energy, Office of Science, Office of Basic Energy Sciences, Division of Chemical Sciences, Geosciences, and Biosciences

\subsection*{Author contribution}
J.Z. conceived the experiments and performed the data analysis. J.Z. and G.D.G. carried out the optical measurements. I. M. carried out the growth of the sample with assistance from G.D.G.  M.T.S. and A.M.D. helped carry out the fridge and optical echo measurements. J.K.B., J.N., and O.G.P. helped carry out the pulsed EPR measurements. All authors contributed to the manuscript.

\subsection*{Competing interests:}
All authors declare they have no competing interests.

\subsection*{Data and materials availability}
All data are available in the main text or the supplementary materials.


\clearpage
\bibliography{references.bib}

\clearpage

\end{document}